\documentclass[%
 reprint,
nofootinbib,
 amsmath,amssymb,
 aps,
]{revtex4-1}

\usepackage{graphicx}% Include figure files
\usepackage{dcolumn}% Align table columns on decimal point
\usepackage{bm}% bold math
 \usepackage{xcolor}

\graphicspath{{./Figures/}}

\begin{document}

\preprint{APS/123-QED}

\title{Robustness of a plasma acceleration based Free Electron Laser}

\author{M. Labat$^{(*)}$}%
	\email{marie.labat@synchrotron-soleil.fr}
\author{A. Loulergue$^{(*)}$}%
\author{T. Andre$^{(*)}$}%
\author{I. Andriyash$^{(*)}$}%
	\email{Present affiliation: Department of Physics of Complex Systems, Weizmann Institute of Science, Rehovot 7610001, Israel}
\author{A. Ghaith$^{(*)}$}%
\author{M. Khojoyan$^{(*)}$}%
	\email{Present affiliation: Laboratoire Leprince Ringuet, UMR 7638, \'Ecole polytechnique, 91 128 Palaiseau Cedex, France}
\author{F. Marteau$^{(*)}$}%
\author{M. Vall\'eau$^{(*)}$}%
\author{F. Briquez$^{(*)}$}%
\author{C. Benabderrahmane$^{(*)}$}%
	\email{Present affiliation: European Synchrotron Radiation Facility, CS 40220, 38043 Grenoble Cedex 9, France}
\author{O. Marcouill\'e$^{(*)}$}%
\author{C. Evain$^{(**)}$}%
\author{M.E. Couprie$^{(*)}$}%
\affiliation{%
 $^{(*)}$Synchrotron SOLEIL, L'Orme des Merisiers, Saint-Aubin, France
}%
\affiliation{
 $^{(**)}$Laboratoire PhLAM, UMR CNRS 8523, Universite Lille 1,
Sciences et Technologies, 59655 Villeneuve d'Ascq, France
}%

\date{\today}% It is always \today, today,
             %  but any date may be explicitly specified

\begin{abstract}
Laser Plasma Accelerators (LPA) can sustain GeV/m accelerating fields offering outstanding new possibilities for compact applications. Despite the impressive recent developments, the LPA beam quality is still significantly lower than in the conventional radio-frequency accelerators, which is an issue in the cases of demanding applications such as Free Electron Lasers (FELs). If the electron beam duration is below few tens of femtosecond keeping pC charges, the mrad level divergence and few percent energy spread are particularly limiting.
Several concepts of transfer line were proposed to mitigate those intrinsic properties targetting undulator radiation applications. 
We study here the robustness of the chromatic matching strategy for FEL amplification at 200~nm in a dedicated transport line, and analyze its sensitivity to several parameters. We consider not only the possible LPA source jitters, but also various realistic defaults of the equipment such as magnetic elements misalignements or focussing strength errors, unperfect undulator fields, etc... 
\begin{description}
%\item[Usage]
%Secondary publications and information retrieval purposes.
\item[PACS numbers]
\pacs{1} 41.60.Cr %May be entered using the \verb+\pacs{#1}+ command.
\pacs{2} 52.38.Kd % Acceleration ->> by laser-plasma interactions
%\item[Structure]
%You may use the \texttt{description} environment to structure your abstract;
%use the optional argument of the \verb+\item+ command to give the category of each item. 
\end{description}
\end{abstract}

\pacs{Valid PACS appear here}% PACS, the Physics and Astronomy
                             % Classification Scheme.
%\keywords{Suggested keywords}%Use showkeys class option if keyword
                              %display desired
\maketitle

\section{Introduction}

Free Electron Lasers (FEL) now provide powerful, tuneable and short pulses in the X-ray spectral range \cite{Ackermann07, Shintake08, Emma10, Ishikawa12, Allaria12NaturePhotonics, Ko_2017, Milne_2017, Wang_2017}. Fourty years after their invention~\cite{Madey71} and first operation in the visible and infra-red range~\cite{Deacon77, Billardon83, warren1983results}, FELs are presently revolutionizing users applications.

As conventional lasers, FELs rely on a gain medium and an amplification process. 
The gain medium consists of free relativistic electrons of Lorentz factor $\gamma$ submitted to the periodic permanent magnetic field of an undulator~\cite{couprie2008xraysCRPhysique}. 
The wiggling of the particles in the magnetic field results in the emission of synchrotron radiation at the so-called "resonance" wavelength $\lambda_r$ given by $\lambda_r = \lambda_u/(2\gamma^2)\times (1+ K_{u}^{2}/2) $ with $\lambda_u$ and $K_{u}$ respectively the undulator period and deflection parameter. The emitted synchrotron radiation (corresponding to the FEL spontaneous emission) further progressing along the undulator, can interact with the electrons resulting in an energy modulation of the electron bunch, which gradually transforms into a density modulation at the resonance wavelength. The emitters are organized in phase, leading to longitudinal coherence and their radiation gets amplified to the detriment of the kinetic energy of the electrons. 
The higher the electronic density, undulator length and resonant wavelength, the higher is the gain. But it falls for lower electron beam energies. Short wavelength FELs thus require long undulators and electron beams of high brightness. 

The FEL wavelength can be tuned by changing the undulator magnetic field or the electron beam energy. In the case of FEL oscillators, for which an optical cavity feeds the radiation back to the undulator gain medium, the tuning range is limited to the VUV~\cite{marsi2002operation} due to gain and mirror performance limitations. Self Amplified Spontaneous Emission (SASE) FELs~\cite{Kondratenko80, Bonifacio84, KimPRL1986}, where the gain is enough to lead to the exponential amplification within one single pass in the undulator, are more suitable for short wavelengths delivery. As a drawback, the SASE FEL pulses longitudinal and spectral distributions present spikes and jitter because the different trains of radiation are not correlated. After low gain coherent harmonic generation~\cite{prazeres1987first}, seeding with an external coherent source tuned at the resonance wavelength was found to dramatically improve the longitudinal coherence and reduce the jitter, intensity fluctuations and gain length~\cite{Yu00HGHG, Lambert08}. Finally, the FEL polarization is determined by the undulator helicity.
X-ray FELs have now reached a high level of maturity, which makes them a unique tool for the exploration of matter~\cite{couprie2008xraysCRPhysique, couprie2015panorama, Bostedt_2016} and which extends the frontiers in terms of light source performance (power, short pulse duration, versatility, etc..). 

Independently, it is also of high interest to explore the possibilities of operating an FEL with emerging technologies. All existing FELs indeed rely on electron beams delivered by conventionnal radio-frequency accelerators. The Laser Plasma Accelerators (LPAs) \cite{tajima:PRL1979} with their recent progresses in terms of beam performance appear to be an attractive alternative and are worth being qualified for an FEL application~\cite{Nakajima08, gruner2007design, couprie2014towards}. 

\begin{figure*}
\includegraphics{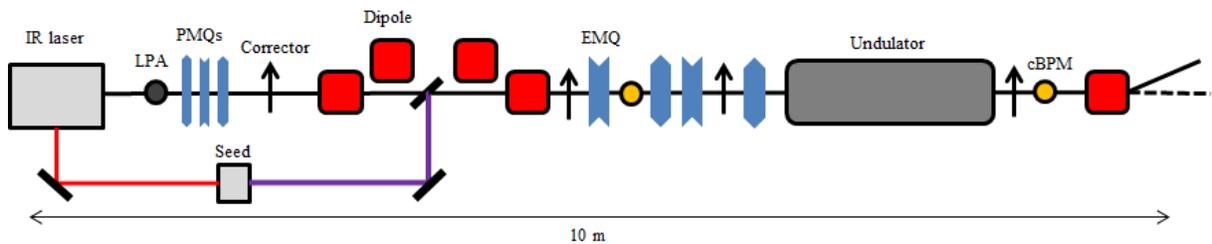}\\
\caption{\label{fig:cox_layout}COXINEL schematic view. LPA: laser plasma accelerator, PMQ: Permanent Magnet Quadrupole, EMQ: ElectroMagnetic Quadrupole, cBPM: cavity Beam Position Monitor~\cite{Hartman-1995,Keil-2013} for orbit correction, Seed: coherent source at 200~nm generated by frequency mixing using the same IR laser as for electron beam generation. Undulator: center located at 6.796~m from the LPA source point.}
\end{figure*}

So far, only synchrotron radiation in the spontaneous regime has been observed on LPAs~\cite{Schlenvoigt08, Fuchs09, Anania09, Lambert12} essentially because the produced electron beams did not fulfill several FEL requirements. The slice energy spread should indeed be smaller than the FEL amplification rate, the normalized emittance $\epsilon_n$ should not be too large with respect to the wavelength $\epsilon_{n}/\gamma < \lambda_r/4 \pi$ and the Rayleigh length should be larger than the gain length \cite{huang2000three}. Besides, initial large divergence and energy spread can induce emittance growth due to chromatic effects~\cite{Migliorati_2013, Floettmann_2003}, which both dramatically reduce the expected gain.   
With an electron beam of typically 1~mrad divergence and 1~$\%$ energy spread, a straightforward amplification cannot be achieved~\cite{couprie2016strategies}. But several strategies can be implemented to overcome those issues.

The electron beam divergence can be handled using plasma lenses~\cite{thaury:SciRep2015} or pulsed plasma lenses~\cite{Nakanii_2015} for LPA beam generation, or implementing strong permanent magnet quadruoples at the LPA exit~\cite{Loulergue-2015}. Concerning the energy spread, several techniques were suggested based for instance on a demixing chicane~\cite{Maier12,couprie2014towards} which sorts the electrons longitudinally according to their energy, or on a transverse gradient undulator \cite{smith1979reducing, Huang12} which sorts them horizontally. For rather long wavelengths, the chicane also enables to stretch the electron beam allowing the optical radiation pulse not to escape from the electron beam longitudinal distribution under the slippage effect (since the photons are travelling slightly faster than the electrons, the radiation pulse naturally progressivly slips ahead of the electron bunch). More recenlty it was also proposed to take advantage of the longitudinal energy sorting introduced by the chicane to synchronize the focussing of the electron bunch slices with the advance of the radiation pulse. In this so-called  chromatic matching regime~\cite{Loulergue-2015}, the effective gain can be significantly increased.

The COXINEL line~\cite{Couprie_PPCF_2016,Couprie_RLE_2017,Andre_NatCom_2018} (see Figure~\ref{fig:cox_layout}) relies on both the implementation of strong permanent magnet quadrupoles at the LPA exit and of the chromatic matching strategy. Following the LPA, a first set of strong permanent magnet quadrupoles enables indeed to refocus the highly divergent electron beam in order to minimize further chromatic effects in the transport line. Four dipoles set in chicane configuration then sort the particles in energy along the longitudinal direction. Following the chicane, four electromagnetic quadrupoles set the chromatic matching optics required for radiation emission and amplification in the last stage, that is the undulator.

In this paper, we present a detailed study of the robustness of the COXINEL transport line enabling FEL amplification in the VUV (200~nm) range. Relying on an LPA reference case compatible with today possiblities, we present the sensitivity of the FEL to the electron beam parameters. We also analyze the dependency of the FEL performance in eventual equipment defaults, taking into account the feasibility of today magnet alignement and magnetic field measurement resolutions.  Finally we discuss perspectives towards shorter wavelengths.

%%%%%%%%%%%%%%%%%%%%%%%%%%%%%%%%%%%%%%%%%%%%%%%%%%%%%%%%%%%%%%%%%%%%%%%%%%%%%%%%%%%%%%%%%%%%
\section{Transport beamline \label{sec:concept}}
%%%%%%%%%%%%%%%%%%%%%%%%%%%%%%%%%%%%%%%%%%%%%%%%%%%%%%%%%%%%%%%%%%%%%%%%%%%%%%%%%%%%%%%%%%%%

\subsection{The LPA electron beam}
The technology of electron beam acceleration in plasma waves has been developing for several decades from first proposals in the late 1970's \cite{tajima:PRL1979} to first experimental demonstrations of narrow spectra beams \cite{faure:Nat04,geddes:Nat04,mangles:Nat04}. In state-of-the-art LPAs, non-linear plasma waves are driven in a gas by femtosecond pulses of a multi-TW laser system, and the bunches of electrons are injected and further accelerated into these waves to high energies ranging from hundreds of MeVs up to few GeVs \cite{leemans:PRL2014}.

The properties of the produced relativistic electron beams depend in a large measure on the injection process. Several possible mechanisms of injection are known and actively discussed in the LPA community. For example, ambient plasma electrons can get injected into the laser driven plasma wave, when the relativistic self-focusing of the laser occurs, rising sharply its intensity and, thus, perturbing the shape of its wake and allowing electrons to get trapped in the accelerating phase of the wake field. This mechanism is known as \emph{self-injection} \cite{bulanov:PRL1997,modena:Nature1995,Kalmykov:PRL2009,corde:NatComm2013}, and it typically results in beam charges up to few hundreds of pico-Coulombs, initial divergence of the order of a few milli-radians, broad energy spreads of tens percents, and with significant shot-to-shot variations (up to few milli-radians). The shot-to-shot stability and angular collimation of LPA electrons can be improved using a mixture of low- and high-Z gases, so that the ionization of high-Z ions occurs in the high laser field and photo-electrons are produced already inside the bubble-like wake structure \cite{McGuffey:PRL2010,Pollock:PRL2011}. The energy spread of the LPA electrons can be reduced down to few percents in the schemes using controlled and rapid injection, e.g. via colliding the driver laser pulse with another one~\cite{esarey:PRL1997,faure:Nat2006} or by introducing a sharp transition of the gas density~\cite{bulanov:PRE1998,geddes:PRL2008,faure:POP2010}.

In experimental conditions, different injection mechanisms can be implemented at the same time, and, in some cases, such hybrid regimes help to improve the overall LPA performance \cite{thaury:SciRep2015}. The energy spread measured on $\sim$~200~MeV electron beams can be as small as 10-20 MeV-fwhm~\cite{thaury:SciRep2015,Pollock:PRL2011}, and their emittance can be below 1~mm.mrad \cite{weingartner:PRSTAB2012}. For such schemes, the full beam charge may vary in the 10 to 100 pC range.

\subsection{Beam optic basics}
The beam dynamics along the manipulation line is here considered, with a specific care on the handling of the intrinsic large divergence and energy spread.

Up to the second chromatic order, the particle coordinates (position and angle in the horizontal and vertical plane respectively) from the source $(x_0,x'_0,z_0,z_0')$ to the undulator center $(x,x',z,z')$ can be presented using the standard transport matrix notation~\cite{Brown}:
\begin{equation}
\label{eq:S2I-x}
\begin{pmatrix}
   x \\
   x' 
\end{pmatrix} = 
\left [
\begin{pmatrix}
   r_{11} & r_{12} \\
   r_{21} & r_{22} 
\end{pmatrix}
+
\delta
\begin{pmatrix}
   r_{116} & r_{126} \\
   r_{216} & r_{226} 
\end{pmatrix}
\right ]
\begin{pmatrix}
   x_0 \\
   x_0' 
\end{pmatrix} 
\end{equation}
\begin{equation}
\label{eq:S2I-z}
\begin{pmatrix}
   z \\
   z' 
\end{pmatrix} = 
\left [
\begin{pmatrix}
   r_{33} & r_{34} \\
   r_{43} & r_{44} 
\end{pmatrix}
+
\delta
\begin{pmatrix}
   r_{336} & r_{346} \\
   r_{436} & r_{446} 
\end{pmatrix}
\right ]
\begin{pmatrix}
   z_0 \\
   z_0' 
\end{pmatrix} 
\end{equation}
with $\delta$ the particle relative energy deviation. 
The first matrix ($r_{ij}$) of the right hand side stands for the linear part and the
second matrix ($r_{ij6}$) stands for the chromatic second order perturbation.

Since the beam exhibits a very large divergence ($\approx$~mrad) and a small transverse size, it may be well approximated by a simple point source with a zero size (or zero emittance). The linear optics can then be simplified to a Source-to-Image (S2I) standard optics, the image being at the center of the undulator.
Indeed, canceling the terms $r_{12}$ and $r_{34}$ respectively in Equations~(\ref{eq:S2I-x}) and (\ref{eq:S2I-z}) enables to set the on-momentum particles ($\delta$ = 0) to a waist $\sigma_{x-min} = r_{11} \sigma_{x0}$ and $\sigma_{z-min} = r_{33} \sigma_{z0}$ as for a standard linear optics imaging the source with magnification $r_{11}$ in the horizontal and $r_{33}$ in the vertical plane. Operating this optics only requires the use of the first quadrupole triplet.

The next step, in the same approximation, is to cancel the $r_{226}$ and $r_{446}$ second order terms at the undulator center, in order to organize the chromatic effects from the large initial divergence. The three rms particle momenta as functions of their relative energy deviation are then approximated by:
\begin{equation}
\label{eq:S2I-rms1}
   \left \{
   \begin{array}{r c l}
      \sigma_{x}^2 (\delta)=r_{11}^2\sigma_{x0}^2 + r_{126}^2 \sigma_{x0}'^2 \delta ^2 \\
      \sigma_{xx'}(\delta)	=r_{126}\sigma'_{x0}\delta/r_{11} \\
      \sigma_{x}'^2(\delta)=\sigma_{x0}'^2/r_{11}^2 \\
   \end{array}
   \right .
\end{equation}
\begin{equation}
\label{eq:S2I-rms2}
   \left \{
      \begin{array}{r c l}
      \sigma_{z}^2 (\delta) =r_{33}^2\sigma_{z0}^2 + r_{346}^2 \sigma_{z0}'^2 \delta^2 \\
      \sigma_{zz'}=r_{346} \sigma_{z0}' \delta / r_{33} \\
      \sigma_{z}'^2(\delta)=\sigma_{z0}'^2/r_{33}^2 \\
   \end{array}
   \right .
\end{equation}
$\sigma_x$, $\sigma'_{x}$ and $\sigma_{xx'}$ are respectively the rms size, divergence and cross term in the horizontal plane (same with $z$ index for the vertical plane).
Because of the chromaticity of this transport, the electron beam is focussed at a different position $S(\delta)$ along the undulator according to:
\begin{equation}
S(\delta) = \ -\ \frac{\sigma_{xx'}(\delta)}{\sigma'^2_{x}(\delta)} = -\ r_{11} r_{126} \delta
\end{equation}

The total geometric emittance $\epsilon_t$ can be derived from the particle momenta integrating over the energy deviation:
\begin{equation}
\label{eq:emitt}
 \left \{
 	 \begin{array}{r c l}
 	\epsilon_{tx}^2 = \epsilon_{x0}^2 + (\frac{r_{126}}{r_{11}}\sigma_{x0}'^2\sigma_{\delta})^2\\
	\epsilon_{tz}^2 = \epsilon_{x0}^2 + (\frac{r_{346}}{r_{33}}\sigma_{z0}'^2\sigma_{\delta})^2\\
   \end{array}
   \right .	
\end{equation}

The second term of the right hand side term corresponds to the chromatic emittance which is added quadratically to the initial one $\epsilon_0$. The chromatic emittance is drastically enhanced by the initial divergence. The normalized emittance, which is most commonly used, simply corresponds to the geometric emittance normalized by $\gamma$.

The chicane strength $r_{56}$ is then used to convert the energy deviation $\delta$ into the longitudinal position $\Delta s$ = $r_{56}\delta$, so that the minimum beam size slips along the bunch all along the undulator.
Since the FEL radiation wave also slips along the bunch at a relative rate of $\lambda_{r}/3\lambda_u$~\cite{Bonifacio90}, the two slippages can be synchronized so that the effective beam size, seen by the FEL, is always minimum. In the exponential gain regime, the synchronization condition can be expressed according to:
\begin{equation}
r_{56} = -\ r_{11} r_{126}\ \frac{\lambda_{r}} {3\lambda_u} 
\end{equation}

This chromatic focussing, which can be seen as a chromatic extension of the linear S2I optics and referred in the following as chromatic matching (S2I-CM) can dramatically improve the FEL process~\cite{Loulergue-2015}. The linear and chromatic tunings are independent of the source and only concern the quadrupole settings. 
To allow a flexible operation of the S2I-CM optics, four quadrupoles, in addition to the first refocussing triplet, were implemented on the COXINEL line.

\subsection{COXINEL magnets}

Since the initial LPA beam divergence tends to dramatically increase the emittance through the first refocussing stage (via $r_{126}$ and $r_{346}$ in Equation~(\ref{eq:emitt}) and (\ref{eq:S2I-rms2})), the first quadrupole triplet is placed as close as possible to the electron beam source point at the cost of very high gradients. 
Permanent magnet quadrupoles of variable strength, designed on purpose, are used~\cite{quapeva_1,quapeva_2,quapeva_3}. They combine a factor 2 gradient tunability, a high maximum gradient and a compact design. 

The demixing chicane relies on four rectangular electromagnetic dipole magnets. Those magnets are designed to ensure a natural closed dispersion and a global straight path, but also to minimize the transverse focussing effects and higher order perturbations. 

Following the chicane, the electron beam matching inside the undulator is ensured by a set of four standard ElectroMagnetic Quadrupoles (EMQs).

To steer the orbit, four correctors are also inserted along the transport line (see Figure~\ref{fig:cox_layout}). They are of window frame type, bipolar and working in both planes with a maximum strength of about 2~mrad.

The space in between the magnetic elements is optimized to fit the mandatory electron beam diagnostics all along the transport line~\cite{Labat_JSR_2018}.
To reach a 200~nm FEL wavelength with a 180~MeV beam, an in-vac+uum undulator~\cite{Marcouille_2015} at its minimum (5.5~mm) gap is used. The generated magnetic field is in the vertical plane.

The final magnetic element is an electromagnetic dipole, similar to the chicane ones, which is used to dump the electron beam away from the photon diagnostics.

%%%%%%%%%%%%%%%%%%%%%%%%%%%%%%%%%%%%%%%%%%%%%%%%%%%%%%%%%%%%%%%%%%%%%%%%%%%%%%%%%%%%%%%%%
\section{Reference case simulations \label{sec:simu-ref}}
%%%%%%%%%%%%%%%%%%%%%%%%%%%%%%%%%%%%%%%%%%%%%%%%%%%%%%%%%%%%%%%%%%%%%%%%%%%%%%%%%%%%%%%%%

%%%%%%%%%%%%%%%%%%%%%%%%%%%%%%%%%%%%%%%%%%%%%%%%%%%%%%%%%%%%%%%%%%%%%%%%%%%%%%%%%%%%%%%%%

\subsection{Reference parameters}

For the sake of simplicity, the simulations do not include the LPA modeling, for which the computationally intense particle-in-cell (PIC) methods should typically be used. For the following studies, an electron beam with gaussian distributions in the six dimensions is assumed, which rms values are listed in Table~\ref{tab:e} for a mean energy of 180~MeV. Since this approach might be slightly optimistic, the (expected small) impact of a more refined initial beam modelling will be investigated in a coming work.
\begin{table}[h]%
\caption{\label{tab:e}%
Electron beam parameters at the LPA source point. Norm. stands for normalized. In the six dimensions, the beam is assumed gaussian and with a cylindrical symmetry along the longitudinal axis.}
\begin{ruledtabular}
\begin{tabular}{lccc}
\textrm{Parameter}& \textrm{Symbol}&  \textrm{Unit} & \textrm{Value}\\
\colrule
Energy & E & MeV & 180 \\
Charge & Q & pC & 34     	 \\
Peak current & $\hat{I}$ & kA & 4  \\
Norm. emittance & $\epsilon$ & $\pi$.mm.mrad & 1  \\
Divergence & $\sigma'_{x}$,$\sigma'_{z}$ & mrad-rms & 1  \\
Energy spread & $\sigma_{\delta}$ & $\%$-rms & 1  \\
\end{tabular}
\end{ruledtabular}
\end{table}

For the further electorn beam transport, all the COXINEL magnets' relevant parameters are detailed in Table~\ref{tab:magnet}.
\begin{table}[h]
\caption{\label{tab:magnet}%
COXINEL magnets' relevant parameters. The three values for PMQs are for each PMQ of the triplet in the beam propagation order.}
\begin{ruledtabular}
\begin{tabular}{lccc}
  & \textrm{Unit}&  \textrm{Value} \\
\colrule
PMQs &&&\\
\colrule
Magnetic length 		& mm 	& 40.7 ; 44.7 ; 26 \\
Min gradient 			& T/m 	& 95.9 ; -97.6 ; 86.7 \\
Max gradient 			& T/m 	& 184.2 ; 186.9 ; 168.6 \\
Gradient for 180~MeV 	& T/m 	& 102.8 ; -103.6 ; 91.9 \\
\colrule
Chicane dipoles &&\\
\colrule
Magnetic length &  mm & 208.33 \\
Integrated field & T.mm & 130 \\
Max $r_{56}$ &  mm & 32 \\
Max field &  T & 0.53 \\
\colrule
EMQs&&\\
\colrule
Magnetic length  & mm & 213.3 \\
Max gradient & T/m & 20 \\
\colrule
Correctors &&\\
\colrule
Max integrated field &  G.m & 38 \\
\colrule
Undulator &&\\
\colrule
Period &  mm & 20 \\
Number of periods & - & 100 \\
Min gap & mm & 5.5 \\
Field at min gap & T & 0.92 \\
\end{tabular}
\end{ruledtabular}
\end{table}

\subsection{Electron beam transport simulation}

The electron beam transport simulations are done in two steps: a first pass with the BETA code~\cite{beta} fitting the first and second order matching to fix the quadrupoles strength, followed by a sympletic 6D tracking pass based on perfect hard edge model magnets~\cite{Forest} from the source to the undulator exit. This simulation tool was benchmarked with ASTRA~\cite{ASTRA} in \cite{Khojoyan_2016}. Transport simulations were also alternatly achieved using OCELOT~\cite{OCELOT}.

As the electron beam dynamics is extremly sensitive to the first quadrupole triplet, a dedicated routine has been written to describe each PMQ gradient profile by slices in the lattice tracking code. The standard technique using the equivalent magnetic length with a constant gradient indeed revealed insufficient. 

\begin{figure}[h]
\includegraphics[width=80mm]{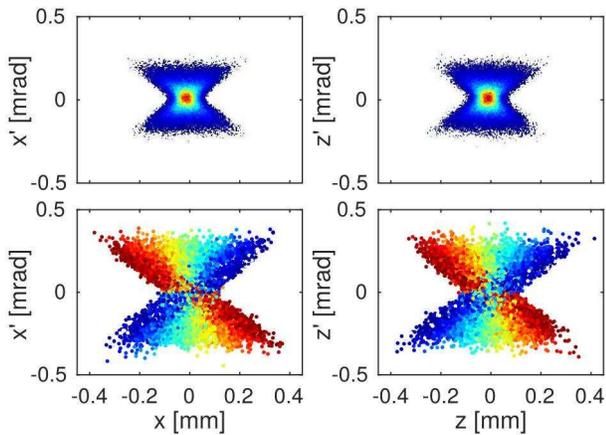} 
\caption{\label{fig:phsp-nonlin-SM} (Left) Horizontal and (right) vertical phase-spaces at the undulator center. (Up) Density plots and (down) colors according to relative energy deviation. Beam transport simulation in the nonlinear case with BETA and sympletic 6D tracking using the S2I-CM optics with $r_{11}$=$r_{33}$=10 and $r_{56}$=0.4~mm.}
\end{figure}

Figure~\ref{fig:phsp-nonlin-SM} shows the electron beam transverse phase-space at the undulator center. According to Equation~\ref{eq:emitt}, the large spanning of the particles results from the large initial divergence and relative energy spread causing chromatic effects along the transport. This spreading requires the use of a modelling including nonlinear dynamics via second order map representation of the transport line components, as it is the case for both codes.

\begin{figure}[h]
\includegraphics{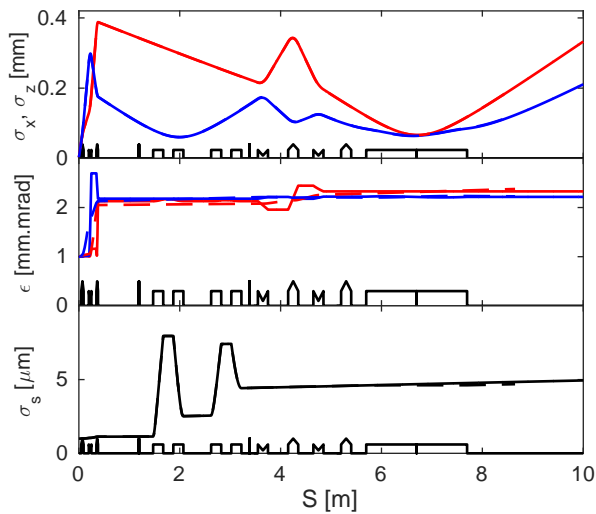}
\caption{\label{fig:env-nonlin-COMP} Beam size envelopes and normalized emittances in the (red) horizontal and (blue) vertical plane, and rms bunch length $\sigma_s$ along the COXINEL line. Beam transport simulation in the nonlinear case with (--) BETA and sympletic 6D tracking and (- - -) OCELOT, both using the S2I-CM optics with $r_{11}$=$r_{33}$=10 and $r_{56}$=0.4~mm.}
\end{figure}

Figure~\ref{fig:env-nonlin-COMP} presents the electron beam size envelopes and emittances along the transport line. The two codes (BETA with sympletic 6D tracking and OCELOT) are found in good agreement. 
The electron beam is focussed at the center of the undulator with a magnification which can be varied from 10 up to 40 (10 in Figure~\ref{fig:env-nonlin-COMP}). The emittance increases from 1 up to about 2.2~$\pi$.mm.mrad in the earlier stage of the focusing due to the strong chromatic effects. Finally, a large transient bunch elongation appears in the chicane due to the horizontal emittance effect. For this low energy (180~MeV) and large energy spread (1$\%$), the different electron speeds also contribute to the bunch lengthening. The linear term over a distance $L$ is simply given by:

\begin{equation}
\label{eq:BL}
r_{56-speed}=L/\gamma^2 
\end{equation}
which gives 0.045~mm at the undulator center. This value is small with respect to the chicane strength region of operation but significantly lengthens the bunch when the chicane is off by about 15$\%$. Higher order terms are negligible here, and this effect can be simply considered as a chicane strength shift.
Another source of bunch lengthening comes from the large divergence of the beam that tends to push backward the large divergence particles. This effect, varying quadratically with the initial divergence, may strongly spoil the benefit of the chicane bunch slice demixing process. Here, this remixing effect is kept relatively low (down to about 15~$\%$ of the initial bunch length for a 1~mrad divergence) but may be drastically increased by larger divergences.

\begin{figure*}
\includegraphics{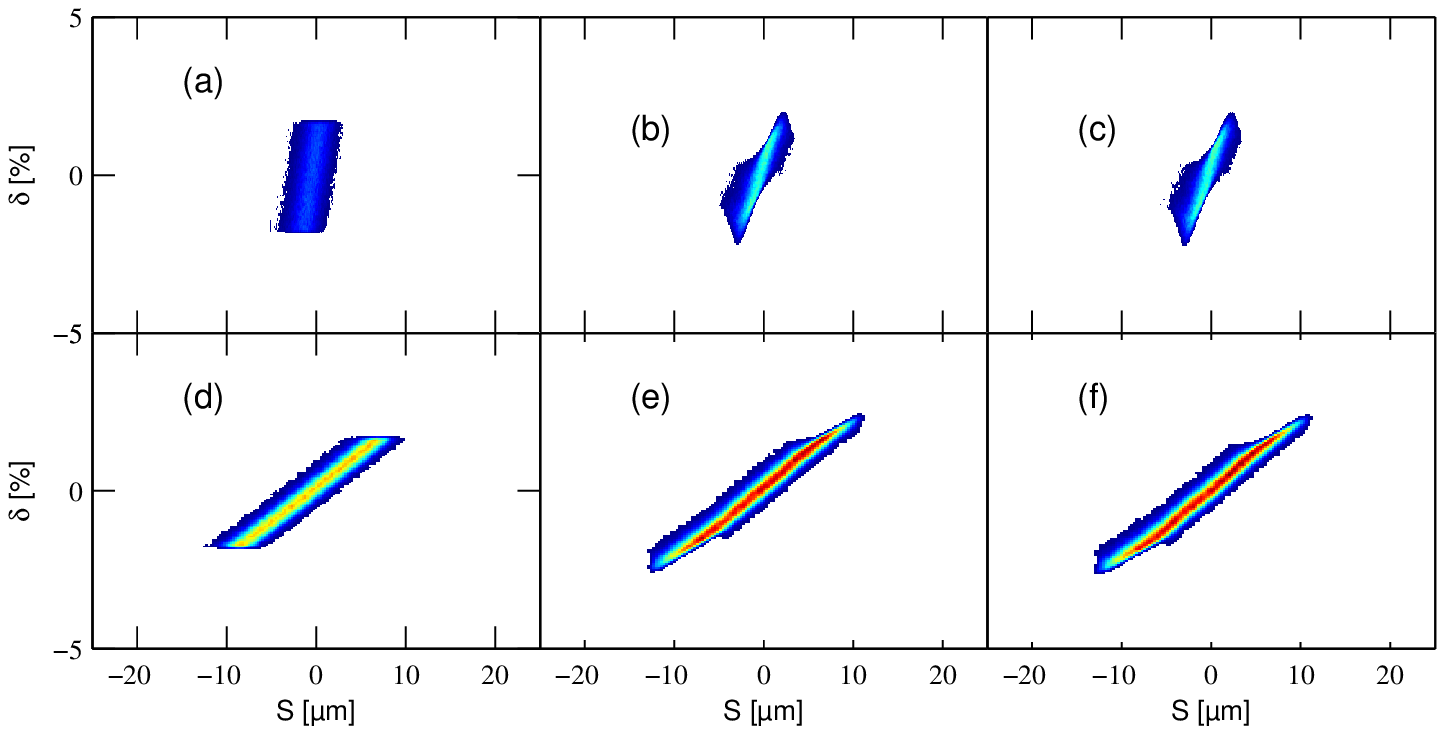}\\
\includegraphics{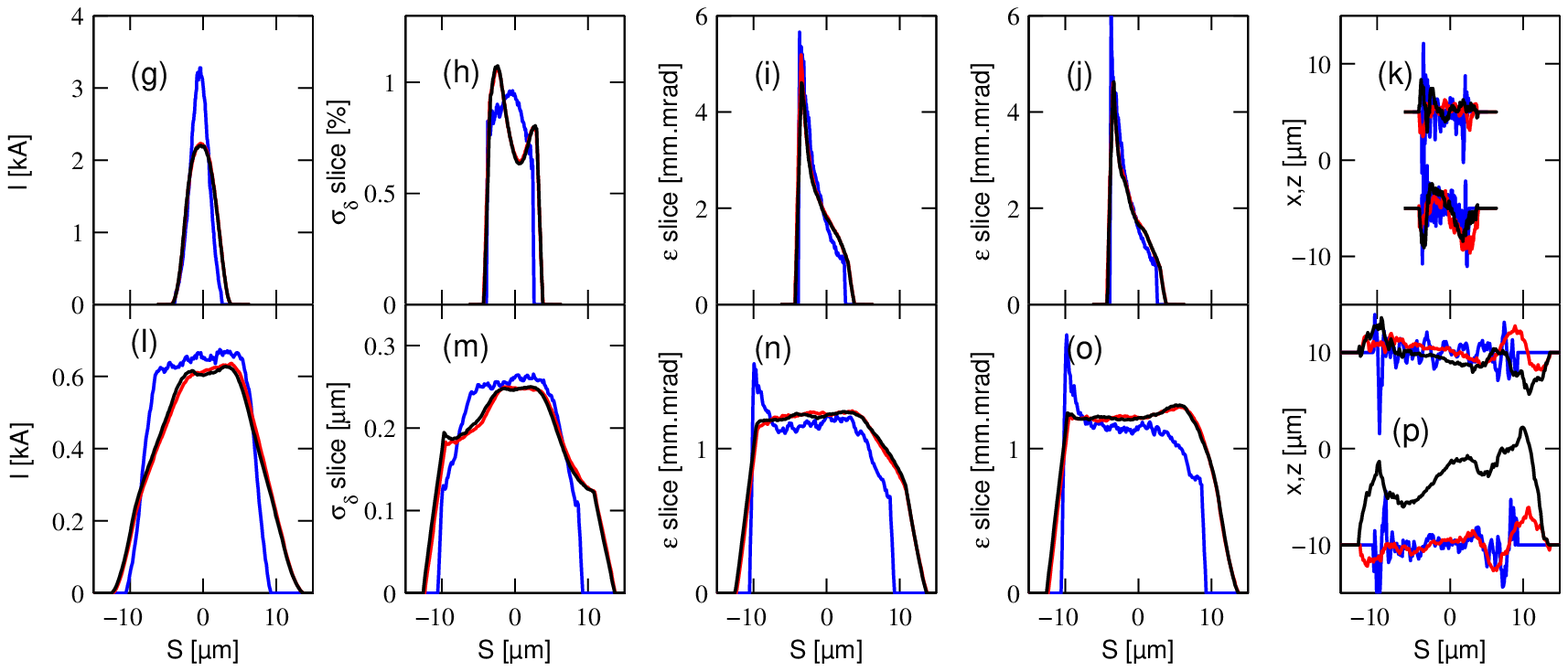}\\
\caption{\label{fig:phsp}Electron beam longitudinal phase-space and corresponding slice parameters at the undulator center. Simulation in the non-linear case: without chicane for (a) and blue curves in (g-k), with chicane for (d) and blue curves in (l-p). Simulation including space-charge forces: without chicane for (b) and red curves in (g-k), with chicane for (e) and red curves in (l-p). Simulation including space-charge forces and CSR effects: without chicane for (c) and black curves of (g-k) and with chicane for (f) and black curves of (l-p). Simulation with BETA and sympletic 6D tracking using the S2I-CM optics with $r_{11}$=$r_{33}$=10 and $r_{56}$=0.4~mm for cases with chicane.}
\end{figure*}

The model can be refined to include collective effects such as space-charge and Coherent Synchrotron Radiation (CSR). The electron beam sensitivity to those effects, without and with chicane, is illustrated in Figure~\ref{fig:phsp}. Without collective effects and without chicane (see Figure~\ref{fig:phsp}(a) and blue lines in (g-k)), the beam is slightly decompressed by the natural particle speed variation according to their energy, and by the trailing particles effect from the initial large divergence. The slice emittances present a large slope due to this same trailing particles effect. With the chicane (see Figure~\ref{fig:phsp}(d) and blue lines in (l-p)), the demixing process is clearly visible on both the energy spread and the emittances at the cost of the peak current drop. 

The space charge can rapidly deteriorate the bunch, especially at low energy and for very short bunches. Indeed, the longitudinal induced energy deviation over a distance $L$ scales as:
\begin{equation}
\label{eq:SC}
\delta \propto \frac{Q}{\gamma^3 \sigma_s^2}L
\end{equation}
where $\sigma_s$ is the rms bunch length and $Q$ is the bunch charge.

Without chicane (see Figure~\ref{fig:phsp}(b) and red lines in (g-k)), the total energy spread is increased by a factor 2, reaching 2$\%$ rms, and enhances the bunch lengthening while reducing the peak current down to 2~kA. The phase-spaces in the three planes are distorted. 
With chicane (see Figure~\ref{fig:phsp}(e) and red lines in (l-p)), phase-space distortions are relaxed, thus at the cost of a lower peak current. Some projected emittance appears, resulting from slice to slice center displacement, but the slice emittance remains below 1.2 $\pi$.mm.mrad while the slice energy spread is almost unchanged.

The chicane decompression relaxes the cumulated space charge effect, but in counter part, adds the CSR effect in the first dipole where bunch length is minimum. Typically, CSR distorts the energy profile all along the bunch inducing some horizontal emittance. In the 1D approximation, the relative energy deviation by longitudinal step $ds$ in a dipole of curvature radius $R$ is given by~\cite{Derbenev}:
\begin{equation}
\label{eq:CSR}
\delta  \propto \frac{Q}{\gamma \sigma_s^{4/3} R^{2/3}} ds
\end{equation}
This effect is mainly enhanced for very short bunch lengths. In the COXINEL configuration, the bunch lengthening in the chicane is dominated by the transient emittance term (thanks to the large chromatic term and initial large divergence) that tends to smooth out the CSR effect especially in the first dipole.
The inclusion of CSR, in addition to space charge, has therefore little impact on the longitudinal beam properties (see (c) and (f), and black lines in (g-p)). It creates a further slice horizontal emittance increase from 1.2 to about 1.3~$\pi$.mm.mrad and the main effect is the slice to slice orbit variations (position and angle) that are large and may suppress the FEL amplification.
Without this transient bunch lengthening, the CSR effect would be more drastic.

%%%%%%%%%%%%%%%%%%%%%%%%%%%%%%%%%%%%%%%%%%%%%%%%%%%%%%%%%%%%%%%%%%%%%%%%%%%%%%%%%%%%%%%%%%%%
\subsection{FEL simulation at 200~nm}
\begin{figure}[h!]
\includegraphics{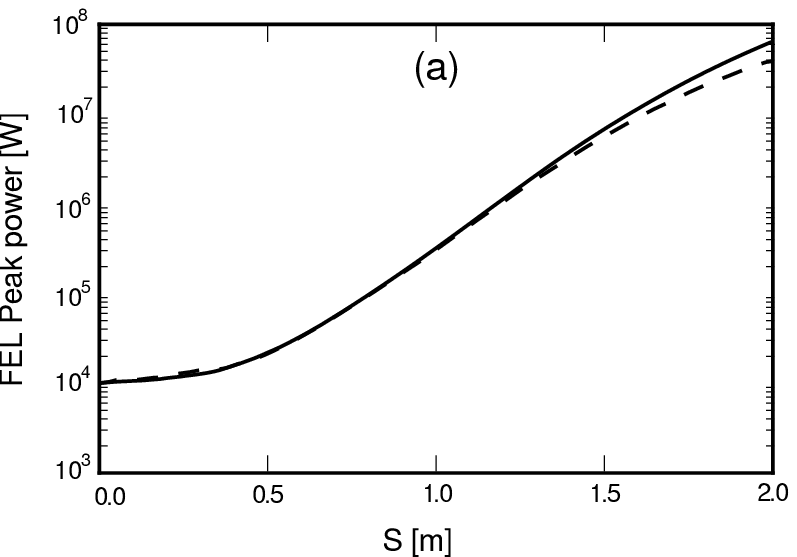}\\
\includegraphics{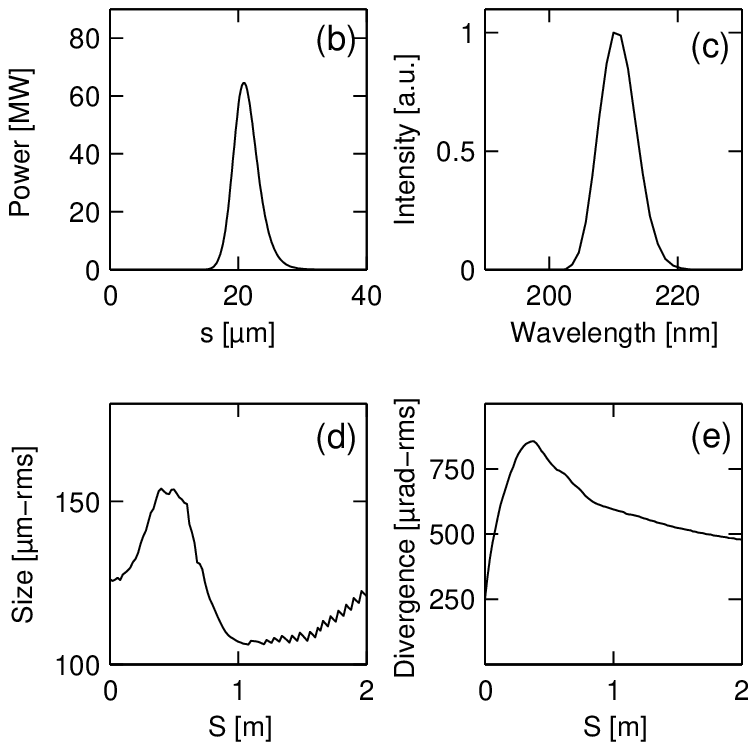}
\caption{\label{fig:fel-200nm} (a) FEL power along the undulator simulated with (--) BETA, sympletic 6D tracking and GENESIS, (- - -) OCELOT and CHIMERA.
(b) FEL peak power longitudinal profile, (c) FEL spectrum at the undulator exit, (d) FEL radiation size and (e) FEL divergence along the undulator, simulated with GENESIS. 
(a-e) Beam transport simulation in the nonlinear case without collective effects and using the S2I-CM optics with $r_{11}$=$r_{33}$=10 and $r_{56}$=0.4~mm.}
\end{figure}

The FEL radiation simulations are performed with GENESIS~\cite{GENESIS} and CHIMERA~\cite{CHIMERA} codes in seeded mode. By default, the seed pulse is defined with a peak power of 10~kW, an infinite pulse duration and a Rayleigh length of 0.5~m with a focussing point at the undulator entrance. The electron beam resulting from the transport simulation described above is used as input while the resonance wavelength is set at 200~nm, for which the U20 undulator gap is set at its minimum (5.5~mm). A linear tapering is applied to the undulator field to compensate the linear energy chirp induced by the chicane. The tapering amplitude is only driven by the magnification parameter, 2.8~$\%$ in the $r_{11}$=$r_{33}$=10 case studied in the following, and remains unchanged while the chicane strength is varied~\cite{Loulergue-2015}. 

\begin{figure}[h!]
\includegraphics{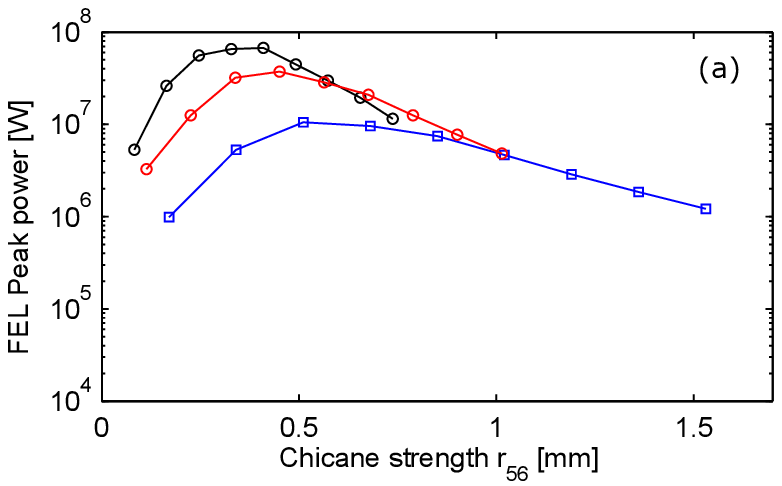}\\
\includegraphics{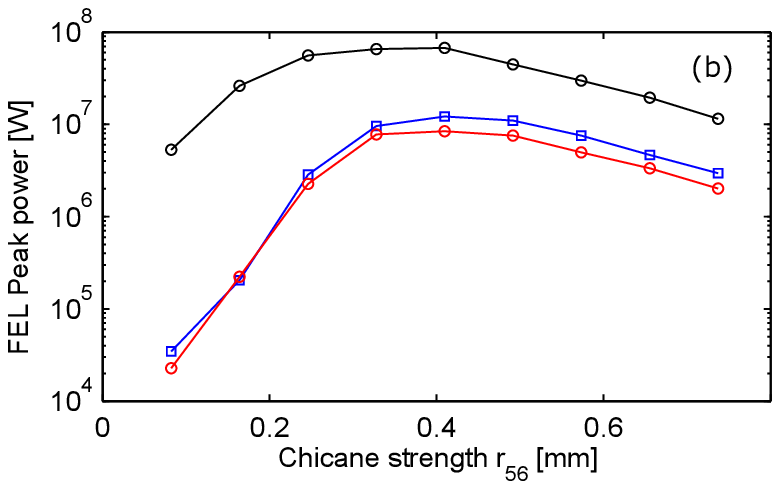}\\
\caption{\label{fig:fel-vs-r56} FEL peak power at 200~nm versus chicane strength. (a) Simulation without collective effects and with magnifications $r_{11}$=$r_{33}$ of (black circle) 10, (red square) 12 and (blue square) 15. (b) Simulation with $r_{11}$=$r_{33}$=10 and (black) without collective effects, (blue) with the space charge and (red) with the space charge and the CSR.
Beam transport simulation in the nonlinear case with BETA and sympletic 6D tracking using the S2I-CM optics. FEL simulation with GENESIS.}
\end{figure}

In the seeded FEL configuration, the initial seed pulse power is exponentially amplified along the undulator following an evolution in $e^{z/L_g}$ where $L_g$ is the so-called gain length. This gain length, directly related to the FEL gain, essentially depends on the electron beam parameters. As presented in Figure~\ref{fig:fel-200nm}(a), both GENESIS and CHIMERA simulations of the FEL power growth along the undulator are found in good agreement. The radiation power reaches $\approx$ 50-60~MW with a gain length of $\approx$ 16~cm (16.3~cm with GENESIS and 16.6~cm with CHIMERA), corresponding to a $\approx$ 0.017 Pierce parameter~\cite{Bonifacio84}. Near the end of the undulator, CHIMERA shows a slightly faster saturation than GENESIS. In the chromatically matched scheme, the saturation is defined by the chromatic focusing dynamics and the radiation slippage. The difference between the codes can therefore be attributed to the fact that the CHIMERA simulation is unaveraged and consequently more sensitive to these phenomena.
The FEL properties at the undulator exit are presented in Figure~\ref{fig:fel-200nm} (b-e). The peak power at the undulator exit is 65~MW (with GENESIS) for a pulse duration of 14~fs-fwhm (4.2$\mu$m-fwhm), corresponding to a pulse energy of 0.9~$\mu$J. The final FEL spectral width is 7~nm-fwhm, i.e. less than 4$\%$, while the spot size is 120~$\mu$m-rms for a divergence below 500~$\mu$rad-rms.

Figure~\ref{fig:fel-vs-r56}(a) presents scans of the FEL output power versus chicane strength $r_{56}$ for different magnifications ($r_{11}$=$r_{33}$).  
The FEL power reached at the optimum $r_{56}$ decreases with the magnification while the optimum $r_{56}$ shifts towards higher values. As expected from~\cite{Loulergue-2015}, the maximum FEL power is reached near the synchronous chicane strength $r_{56-synch}$.

Figure~\ref{fig:fel-vs-r56}(b) then depicts the influence of the collective effects on the FEL performance. The space charge typically spoils the peak power by one order of magnitude. In these investigations, the quadrupole strengths are not re-tuned.

%%%%%%%%%%%%%%%%%%%%%%%%%%%%%%%%%%%%%%%%%%%%%%%%%%%%%%%%%%%%%%%%%%%%%%%%%%%%%%%%%%%%%%%%%%%%
\section{Sensitivity to LPA parameters \label{sec:simu-lpa}}
%%%%%%%%%%%%%%%%%%%%%%%%%%%%%%%%%%%%%%%%%%%%%%%%%%%%%%%%%%%%%%%%%%%%%%%%%%%%%%%%%%%%%%%%%%%%
In order to evaluate the robustness and the main dependencies of the expected lasing effect, the sensitivity to the LPA main parameters and inherent jitters is first studied.

\subsection{Sensitivity to LPA charge and divergence}

\begin{figure}[h!]
\includegraphics[width=80mm]{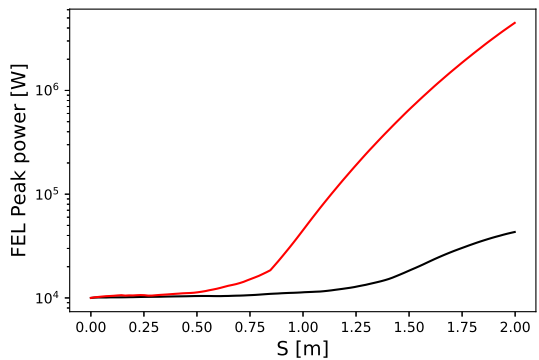}
\caption{\label{fig:fel_vs_charge} FEL peak power at 200~nm along the undulator with a (black) 1~pC and (red) 5~pC per $\%$ of energy bandwidth. Beam transport simulation with GENESIS without collective effects using the S2I-CM optics with $r_{11}$=$r_{33}$=10 and $r_{56}$=0.4~mm. Normalized emittance: 0.2~mm.mrad, divergence: 1~mrad. FEL simulation with GENESIS.}
\end{figure}

The most critical LPA parameters for the final FEL performance remain the charge and the divergence. Whatever the efforts on the further beam transport, a minimum initial charge per energy bandwidth together with a reasonable initial divergence are mandatory for an amplification to occur. As illustrated for instance in Figure~\ref{fig:fel_vs_charge}, an initial charge of 1~pC per $\%$ of energy bandwidth hardly enables to enhance the input seed, while with 5~pC per $\%$, the amplification reaches more than two orders of magnitude. 

\begin{figure}[h!]
\includegraphics[width=100mm]{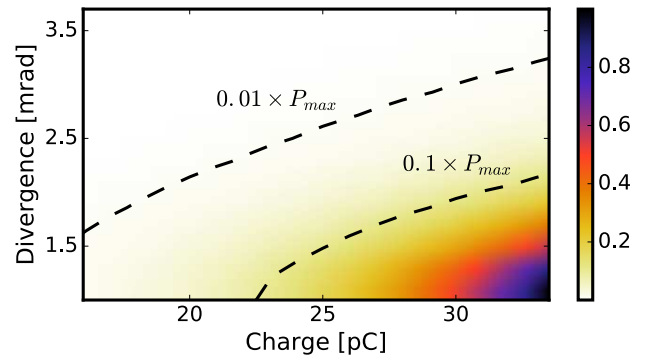}
\caption{\label{fig:fel_scan} Normalized FEL peak power at 200~nm versus electron beam charge and divergence. Beam transport simulation with OCELOT without collective effects using the S2I-CM optics with $r_{11}$=$r_{33}$=10 and $r_{56}$=0.4~mm. Normalized emittance preserved at 1~mm.mrad. FEL simulation with CHIMERA.}
\end{figure}

Following, Figure~\ref{fig:fel_scan} displays in 2D the FEL peak power dependency to the LPA beam charge and divergence. Starting from the reference case with 34~pC charge and 1~mrad-rms divergence, the peak power drops by one order of magnitude if the charge is reduced by 10~pC or the divergence increased by 1~mrad.  For charges below 15~pC or divergences above 3~mrad, the power decrease is found of two orders of magnitude. The FEL performance is extremely sensitive to these fundamental parameters.

\subsection{Sensitivity to LPA jitters}

The initial source jitters have to be considered as inherent jitters and are uncorrelated to the laser pointing. In the following, they are assumed to be of the order of 5~$\mu$m-rms in position, 1~mrad-rms in angle and 1~$\%$-rms in energy spread. 
\begin{figure}[h!]
\includegraphics{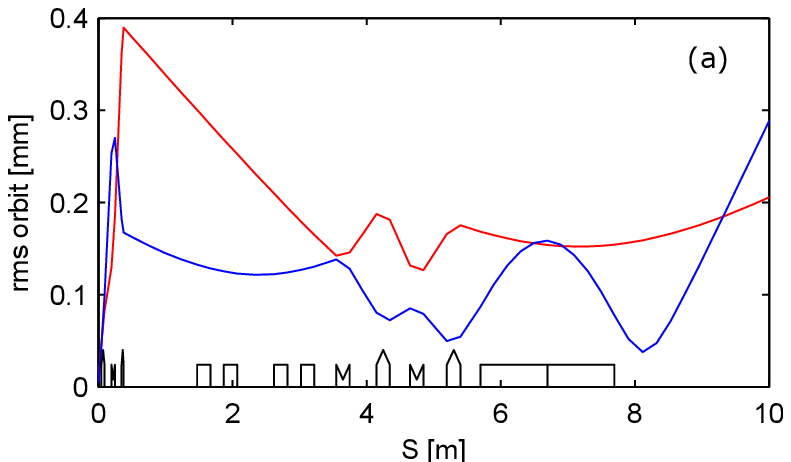}\\
\includegraphics[angle=90]{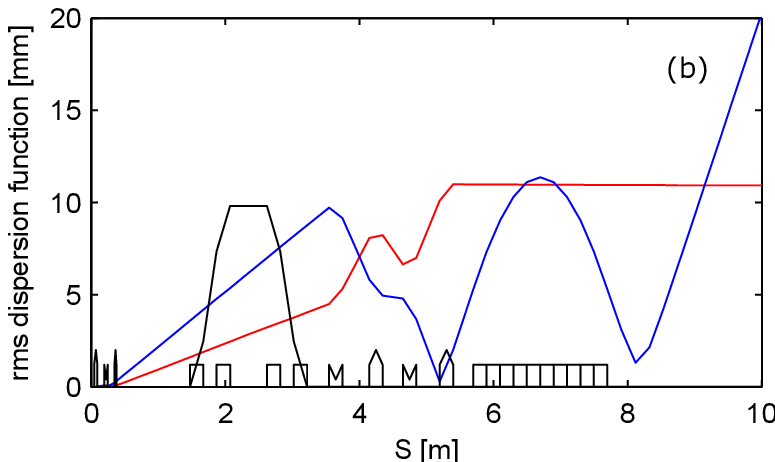}\\
\includegraphics{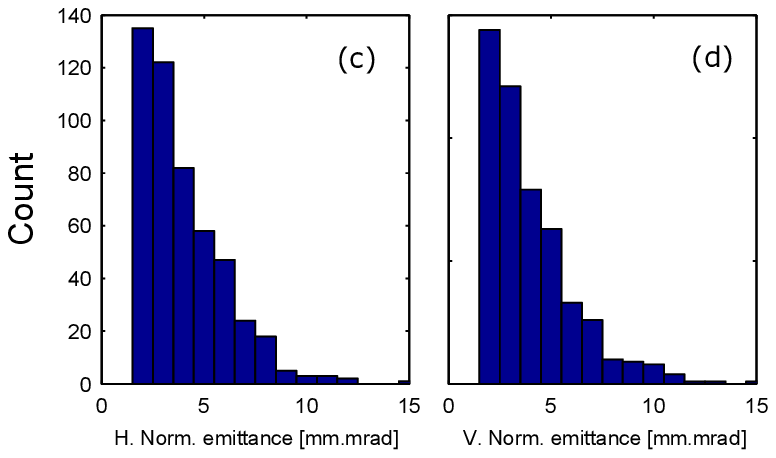}\\
\caption{\label{fig:Q-vs-lpa} Effect of random source jitters (5~$\mu$m-rms in position, 1~mrad-rms in divergence and 1~$\%$-rms in energy spread) on the LPA beam quality. (a) Rms orbits in the (red) horizontal and (blue) vertical plane.
(b) Dispersion functions in the (red) horizontal plane, (blue) vertical plane and (black) horizontal plane without jitter.
(c-d) Statistical emittance distribution (500 tries) without the energy spread jitter, but with position and divergence jitters (5~$\mu$m-rms in position, 1~mrad-rms in divergence) and including orbit corrections. 
%The initial unpertubed emittances are about 2 $\pi$.mm.mrad in both planes.
Simulation in the nonlinear case with BETA and sympletic 6D tracking without collective effects using the S2I-CM optics with $r_{11}$=$r_{33}$=10 and $r_{56}$=0.4~mm.}
\end{figure}

Figure~\ref{fig:Q-vs-lpa}(a) displays the rms orbits induced by those jitters. Thanks to the S2I-CM optics, the orbits are limited to about 0.15~mm and 0.15~mrad in the undulator and are dominated by the 1~mrad-rms pointing. 

Another risk of source jitters is the generation of both horizontal and vertical spurious dispersion functions when passing off-axis through the quadrupoles. Small quadrupole offsets act as a dipole spanning the particle trajectories further amplified by the focusing of the quadrupoles according to their energy. Even with the orbits corrected, some dispersion persists. 
The dispersion functions are plotted in Figure~\ref{fig:Q-vs-lpa}(b) and exhibit values up to 10~mm in the undulator region, i.e. as large as the purposed chicane dispersion function. This effect spreads the beam up to a final size of 100~$\mu$m in the case of 1$\%$ energy spread. This value is equivalent or larger than the beam waist size in the undulator and may spoil the FEL efficiency.

An immediate consequence of the dispersion increase, even after orbit correction, is the total emittance growth by center of mass displacement for each energy slice. Figure~\ref{fig:Q-vs-lpa}(c-d) presents the emittance distribution based on random source jitters (5~$\mu$m-rms and 1 mrad-rms) with orbit correction. The total emittances are typically increased by a factor of 3 to 5 from the pure chromatic focusing dispersion.

\begin{figure}[h!]
\includegraphics{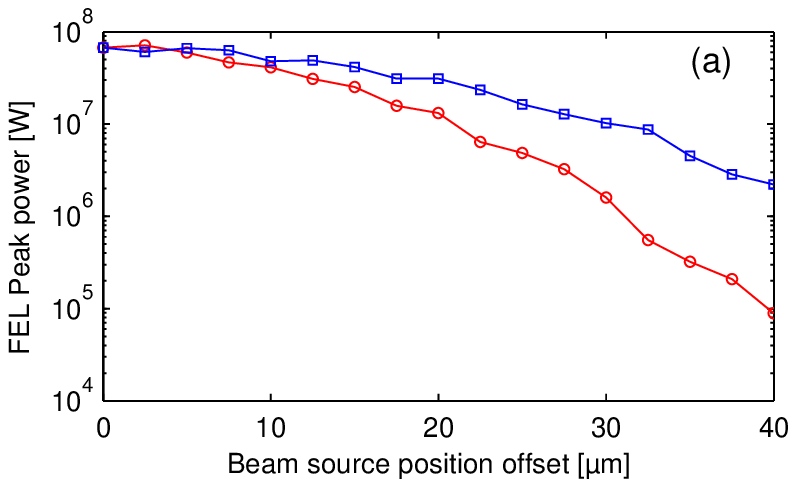}\\ 
\includegraphics{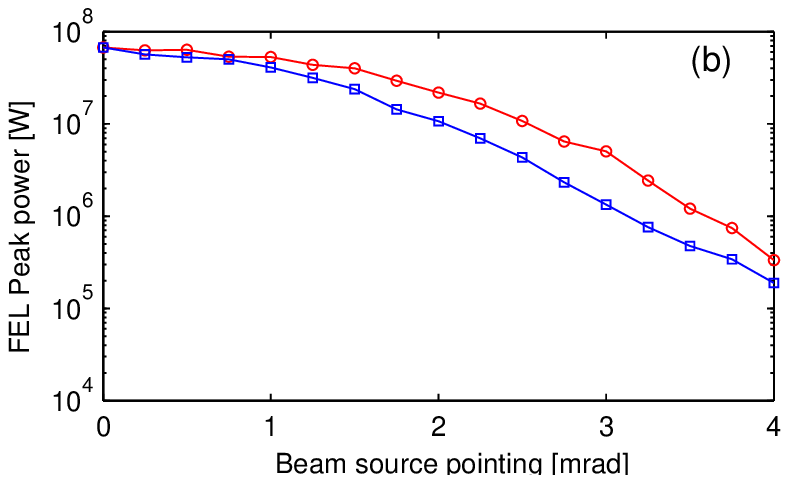}\\ 
\caption{\label{fig:fel-vs-lpa-1} Normalized FEL peak power at the undulator exit versus initial LPA beam (a) position and (b) pointing (angle) in the (red circle) horizontal and (blue square) vertical plane. Beam transport simulation in the nonlinear case with BETA and sympletic 6D tracking without collective effects using the S2I-CM optics with $r_{11}$=$r_{33}$=10 and $r_{56}$=0.4~mm. FEL simulation with GENESIS.}
\end{figure}

Figure~\ref{fig:fel-vs-lpa-1} shows the impact of the possible LPA position and pointing jitters on the FEL output power. A drop by one order of magnitude is reached for displacements above 20~$\mu$m-rms and angles above 2~mrad-rms. The FEL power decreases for large beam position offsets and pointings mainly results from the loss of overlap between the electron beam and the external seed which remains injected on-axis.

\subsection{Electron beam loss sensitivity}

Another important consequence of large LPA beam parameter deviations from the reference case can be an increase of the beam losses along the transport line, at the risk of permanent magnets (PMQs and undulator) demagnetization. 

\begin{figure}
\includegraphics[width=80mm]{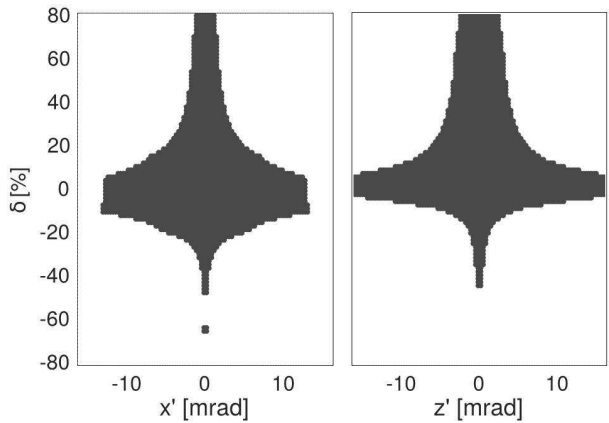}
\caption{\label{fig:acceptance} (Left) horizontal and (Right) vertical acceptance of the transport line. Beam energy at 180~MeV using S2I-CM optics with $r_{11}$=$r_{33}$=10 and $r_{56}$=0.4~mm. Beam transport simulation in the nonlinear case with BETA and sympletic 6D tracking.}
\end{figure}

\begin{figure}
\includegraphics[width=80mm]{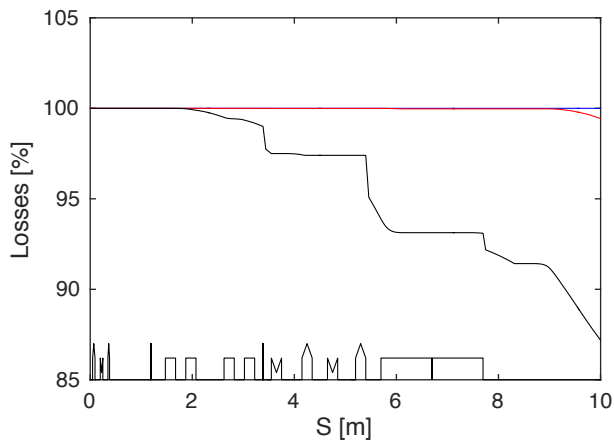}
\caption{\label{fig:losses} Beam losses along COXINEL line assuming an energy spread of (blue) 1~$\%$-rms, (red) 10~$\%$-rms and (black) 20~$\%$-rms together with a divergence of 1~mrad-rms. Beam transport simulation in the nonlinear case with BETA and sympletic 6D tracking without collective effects using the S2I-CM optics with $r_{11}$=$r_{33}$=10 and $r_{56}$=0.4~mm.}
\end{figure}

The vacuum chamber diameter along the transport line is essentially 10~mm, but it falls down to less than 7~mm in the three PMQs, the cBPMs and in the undulator in the vertical plane when the gap is closed to 5.5~mm. 
Assuming this geometry, the final acceptance as a function of the LPA pointing and beam energy spread at the source point is illustrated in Figure~\ref{fig:acceptance}. The dark area corresponds to the space in which the particles can be transported down to the undulator exit. Particles in the white area are lost in the vacuum chambers before reaching the undulator exit. 
The acceptance falls below the +/-20~$\%$ level already above a few mrad of pointing, which would correspond to a drastic rise of the beam losses. The main initial source parameters that drive large radial excursions are the divergence and the energy deviation. The losses are therefore estimated for three different levels of energy spread $\delta$ using an initial divergence of 1~mrad (see Figure~\ref{fig:losses}). For $\delta$ below 10~$\%$, the losses are negligible. But above 20~$\%$ less than 90~$\%$ of the initial charge reaches the end of the line.

%%%%%%%%%%%%%%%%%%%%%%%%%%%%%%%%%%%%%%%%%%%%%%%%%%%%%%%%%%%%%%%%%%%%%%%%%%%%%%%%%%%%%%%%
\section{Sensitivity to equipment defects \label{sec:simu-equip}}
%%%%%%%%%%%%%%%%%%%%%%%%%%%%%%%%%%%%%%%%%%%%%%%%%%%%%%%%%%%%%%%%%%%%%%%%%%%%%%%%%%%%%%%%
In addition to the inherent LPA source jitters, misalignments and/or defaults of the magnetic components can also impressively affect the expected FEL performance.

\subsection{Electron beam sensitivity}

To study the electron beam sensitivity to those effects, a random displacement of all the quadrupoles magnetic center by up to 100~$\mu$m, together with a random tilt of the dipoles of up to 100~$\mu$m and a random error on their possible relative strength of up to 0.1$\%$, are applied to the transport line simulation.

\begin{figure}[h]
\includegraphics{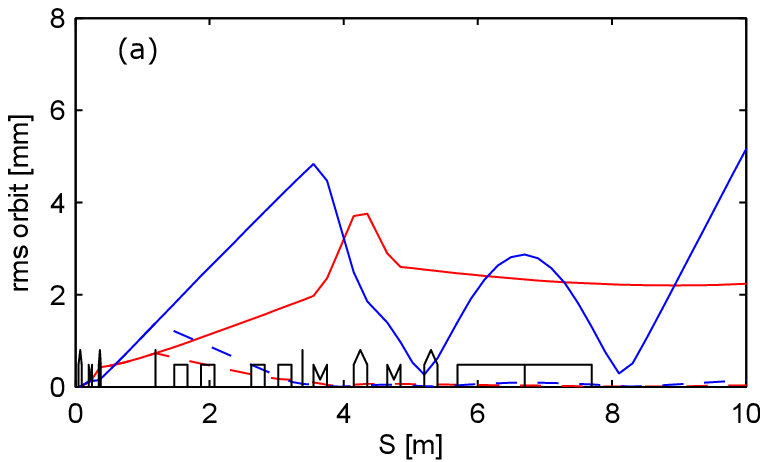}\\
\includegraphics[angle=90]{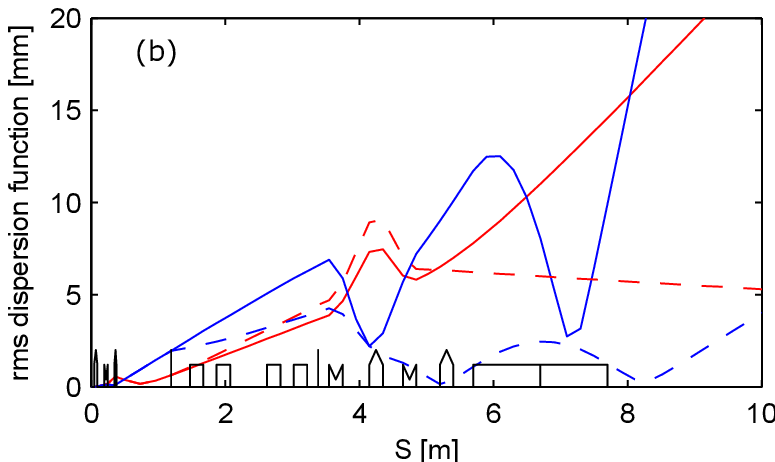}\\
\includegraphics{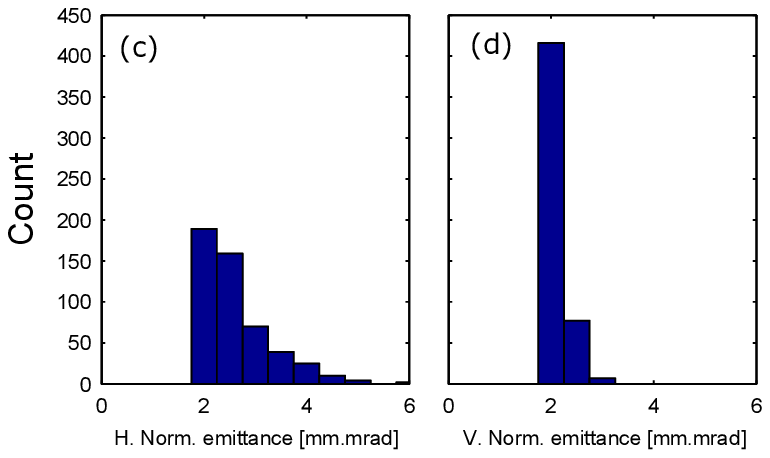}\\
\caption{\label{fig:beam-equip} Effect of random magnet misalignments/defaults on the electron beam (a) rms orbits and (b)rms dispersion functions, in the (red) horizontal and (blue) vertical plane, (--) without and (- - -) with orbit correction. (c-d) Effect on the statistical normalized emittance distribution in the horizontal and vertical plane (500 tries) with orbit correction. Initial undisturbed emittances are about 2 $\pi$.mm.mrad in both planes. (a-b) Random displacement of all the quadrupoles magnetic center by up to 100~$\mu$m, together with a random tilt of the dipoles of up to 1~mrad and a random error on their possible relative strength of up to 0.1$\%$. (c-d) Random displacement of all the quadrupoles only.
Beam transport simulation in the nonlinear case with BETA and sympletic 6D tracking without collective effects using the S2I-CM optics with $r_{11}$=$r_{33}$=10 and $r_{56}$=0.4~mm.}
\end{figure}

Figure~\ref{fig:beam-equip}(a) displays the effect on the rms orbits without and with orbit correction. Without orbit correction, up to 5~mrad slopes together with 5~mm-rms orbit amplitudes are obtained in the undulator, i.e. quite above the half gap aperture of the undulator. Once the correction is applied using the two first correctors on the two cBPMs, the orbits are strongly reduced down to 0.4~mm in the undulator. 

Figure~\ref{fig:beam-equip}(b) displays in turn the effect on the rms spurious dispersion functions, again without and with orbit correction. The dispersion functions are large at the undulator location (about 10~mm) and almost unaffected by the orbit correction. This means that even with a very small orbit measured on both cBPMs, the spurious dispersion functions still remain and may strongly affect the FEL amplification.

The created dispersion functions rise an emittance increase by center of mass displacement just as in the case of source jitter (see section~\ref{sec:simu-lpa}). 
As shown in Figure~\ref{fig:beam-equip}(c), including only the quadrupoles magnetic center random displacement and a further orbit correction, the total emittances typically increase by a factor of 3 to 5 from the pure chromatic focusing dispersion.

Those large orbit deviations and dispersion functions are usually corrected on conventionnal accelerators using Beam Based Alignment techniques~\cite{Emma_1999}. In the specific framework of LPA, a specific beam pointing alignment compensation (BPAC) strategy can be applied~\cite{Andre_NatCom_2018}. Taking advantage of the motorised translations of the PMQs while monitoring the electron beam transverse shape on dedicated diagnostics, the quadrupole magnetic centers are tuned to independently minimise the transverse offsets as well as the dispersion functions. This method supresses the eventual misalignments of the quadrupoles as well as of the LPA laser.

\subsection{FEL sensitivity}

\subsubsection{Sensitivity to quadrupole offset}

\begin{figure}
\includegraphics{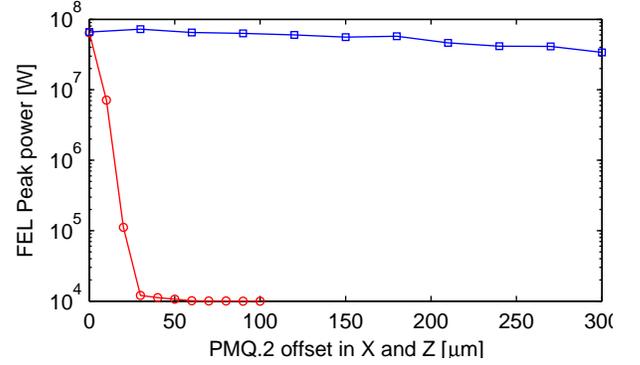} 
\caption{\label{fig:fel-200nm-vs-PMQoffset} FEL peak power at 200~nm at the undulator exit versus second PMQ radial offset (blue) with and (red) without orbit correction. Beam transport simulation in the nonlinear case with BETA and sympletic 6D tracking without collective effects using S2I-CM optics with $r_{11}$=$r_{33}$=10 and $r_{56}$=0.4~mm. FEL simulation with GENESIS.}
\end{figure}

If neither BPAC or orbit correction are applied, systematic errors on magnet alignment will spoil the final FEL gain via, as previously detailed, orbit distorsion, dispersion functions and emittance increases. As shown in Figure~\ref{fig:fel-200nm-vs-PMQoffset} for instance, an offset of 10~$\mu$m on the second PMQ, would lead to a drop of the FEL output power by one order of magnitude. The performance degradation here essentially comes from the impact of the PMQ displacement on the orbit. The electron beam transverse size in the undulator is indeed around 80~$\mu$m while the seed's is 200~$\mu$m and the FEL radiation's between 100 and 150~$\mu$m. Without orbit correction, a 30~$\mu$m offset of the second PMQ induces an orbit displacement of 400~$\mu$m driving the electron beam out of the seed path.

With orbit correction (or BPAC), the power drop is limited to negligible values up to 200~$\mu$m displacements before loosing a factor 2 at 300~$\mu$m. In this case, the FEL power decrease results from  the second order effect of the dispersion. For a 300~$\mu$m offset of the second PMQ, the dispersion function rises up to 8~mm inducing an orbit displacement of 80~$\mu$m which drives the electron beam in low power regions of the seed.

\subsubsection{Sensitivity to quadrupole strength}

\begin{figure}
\includegraphics{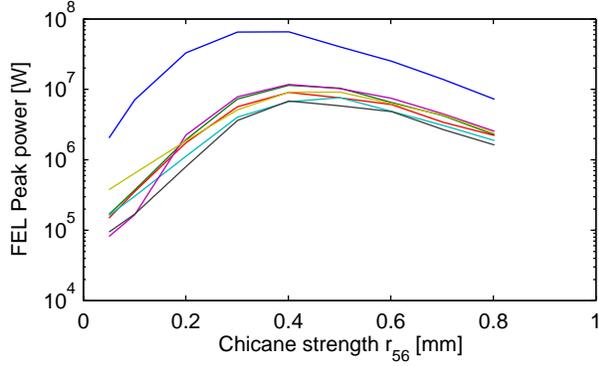}\\
\caption{\label{fig:fel-vs-r56-200nm-wQdefo} FEL peak power at the undulator exit versus chicane strength for various quadrupole strength shifts: (blue) ideal, (dark green) +8$\%$ on PMQ 1, (red) +1$\%$ on PMQ 2, (light blue) +1$\%$ on PMQ 3, (purple) +8$\%$ on EMQ 1, (light green) +4$\%$ on EMQ 2, (black) +20$\%$ on EMQ 3. Beam transport simulation in the nonlinear case with BETA and sympletic 6D tracking without collective effects using the S2I-CM optics with $r_{11}$=$r_{33}$=10. FEL simulation with GENESIS.}
\end{figure}

In Figure~\ref{fig:fel-vs-r56-200nm-wQdefo}, the FEL performance is scanned versus the chicane strength for various quadrupole strength shifts. In order to classify the individual quadrupole level of tuning sensitivity, we detuned their strength one by one until a drop in the FEL output power by one order of magnitude is observed. 
Depending on the considered quadrupole, the one order of magnitude drop is reached for errors in the 1 to 20$\%$ range. The most dramatic effect results from an error on the second PMQ having the largest integrated strength together with large optical functions.
Since with the S2I-CM optics the beam focussing is highly chromatic in the undulator, a detuning of the PMQs and/or EMQs with respect to the setting for the nominal energy leads to severe mismatches in the undulator and therefore to a significative degradation of the FEL gain.

\subsubsection{Sensitivity to the undulator magnetic field}
\begin{figure}
\includegraphics{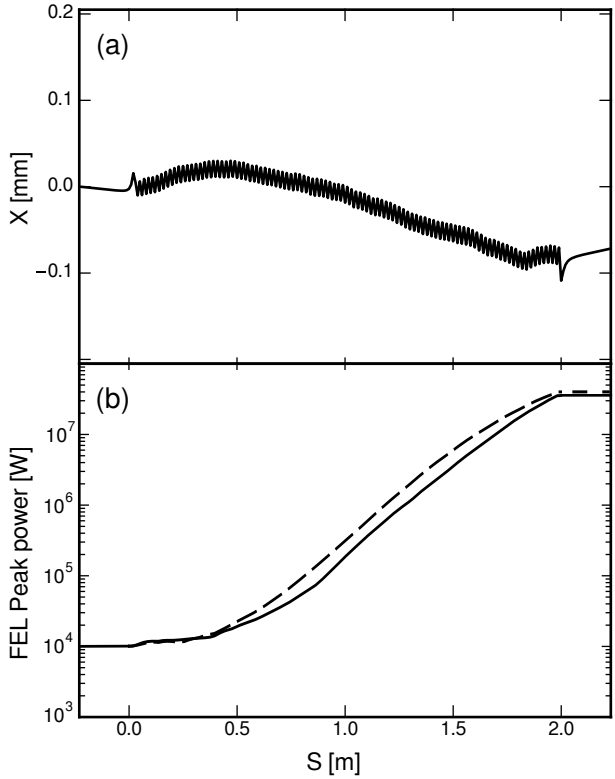}\\
\caption{\label{fig:FEL-vs-B-und}(a) Electron beam trajectory in the horizontal plane using measured undulator magnetic field and (b) FEL power along the undulator using (--) measured and (- - -) purely sinusoidal undulator magnetic field. Beam transport in the nonlinear case with OCELOT using the S2I-CM optics with $r_{11}$=$r_{33}$=10 and $r_{56}$=0.4~mm. FEL simulation with CHIMERA.}
\end{figure}

The CHIMERA unaveraged FEL simulations enable to study the effect of a non purely sinusoidal, i.e. real, undulator magnetic field, where the small phase and amplitude errors may reduce the effective gain. For the real case simulation, the magnetic field measurements of an existing U20 undulator of SOLEIL~\cite{couprie2010panoply} are used, while in the ideal case, the magnetic field is perfectly sinusoidal without any phase or amplitude errors.
In the measured undulator case, electrons with moderate energies may deviate significantly from the axis, essentially in the horizontal plane, as shown in Figure~\ref{fig:FEL-vs-B-und}(a). To find the best correction, series of FEL simulations were considered varying the beam entrance angle and an optimum was found for a small angle of 27 $\mu$rad. 
After optimization of the undulator taper in both cases, the FEL power growths along the undulator are compared in Figure~\ref{fig:FEL-vs-B-und}(b). 
The final powers are rather close though non-negligible differences in the amplification dynamics can be observed. In the measured undulator case, the amplification includes the part from 50 cm to 90 cm where the gain is partially suppressed by the beam off-axis drift. In addition, the pointing at the undulator exit is found slightly off-axis. 
This study reveals that the imperfections of the undulator magnetic field can easily be overcome using the appropriate orbit correction at the undulator entrance and exit.

\subsubsection{Sensitivity to undulator taper}

\begin{figure}[h]
\includegraphics{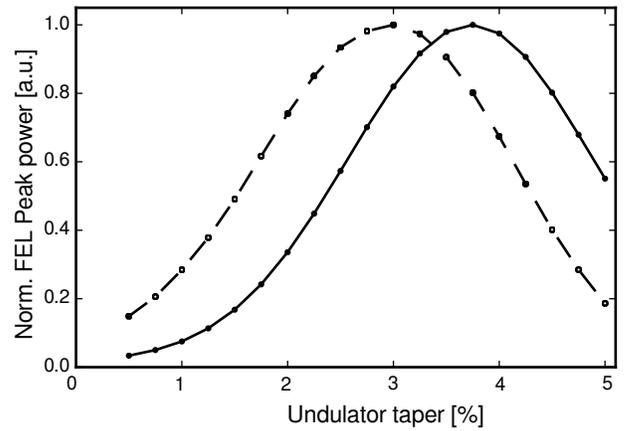}
\caption{\label{fig:fel-200nm-vs-taper} FEL peak power versus undulator taper using (--$\bullet$--) real and (- $\square$ -) ideal undulator magnetic field. Beam transport in the nonlinear case with OCELOT using the S2I-CM optics with $r_{11}$=$r_{33}$=10 and $r_{56}$=0.4~mm. FEL simulation with CHIMERA.}
\end{figure}

The sensitivity of the FEL output peak power to the undulator taper is illustrated in Figure~\ref{fig:fel-200nm-vs-taper} in both the measured and ideal undulator cases presented previously. Their behaviours are rather similar and a fall of one order of magnitude in the FEL output power is reached for taper errors of less than 3~$\%$. This level of precision in the tapering is mechanically achievable on standard U20 undulators in operation at SOLEIL (less than 10~$\mu$ precision achieved).

%%%%%%%%%%%%%%%%%%%%%%%%%%%%%%%%%%%%%%%%%%%%%%%%%%%%%%%%%%%%%%%%%%%%%%%%%%%%%%%%%%%%%%%%
\section{Perspectives at shorter wavelengths}
%%%%%%%%%%%%%%%%%%%%%%%%%%%%%%%%%%%%%%%%%%%%%%%%%%%%%%%%%%%%%%%%%%%%%%%%%%%%%%%%%%%%%%%%

The sensitivity and robustness of the LPA-based FEL has been examined first in the 200~nm case. An extension to a lower wavelength, 40~nm, is possible using a 400 MeV beam. 
The beam transport layout is exactly the same as in the 200~nm case, except that the initial PMQs setting is changed accordingly to the energy increase and that the U20 undulator is replaced by a further smaller period U15 undulator. The COXINEL magnets configuration for this case is summarized in Table~\ref{tab:magnet_40nm}. The required seeding power remains 10~kW at 40~nm which is easily achievable using a high order harmonics generated in gas.
\begin{table}
\caption{\label{tab:magnet_40nm}%
COXINEL magnets' relevant parameter for the 40~nm wavelength case using a 400~MeV beam. The three values for PMQs are for each PMQ of the triplet in the beam propagation order.}
\begin{ruledtabular}
\begin{tabular}{lccc}
  & &  \textrm{Values} \\
\colrule
PMQs &&&\\
\colrule
Magnetic length [mm] & 66 & 81.1 & 47.1 \\
Min gradient [T/m] & 103.6 & -106.2 & 98.5 \\
Max gradient [T/m] & 195.8 & 199.6 & 188.2 \\
Gradient for 400~MeV [T/m] & 142.2 & -142.5 & 133.4 \\
\colrule
Undulator &&\\
\colrule
Period [mm]& & 15 \\
Number of periods && 200 \\
Nominal gap [mm] & & 3.5 \\
Field at nominal gap [T] & & 1.52 \\
\end{tabular}
\end{ruledtabular}
\end{table}

\begin{figure}
\includegraphics{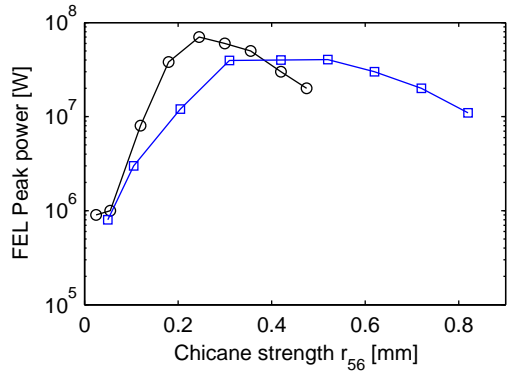}
\caption{\label{fig:fel-vs-r56-40nm} FEL peak power at 40~nm at the undulator exit versus chicane strength using a magnification of (black circle) 15 and (blue square) 20. Beam transport simulation in the nonlinear case with BETA and sympletic 6D tracking without collective effects using the S2I-CM optics with a 400~MeV beam. FEL simulation with GENESIS.}
\end{figure}

The peak FEL power reached at 40~nm versus the chicane strength for two magnifications is presented in Figure~\ref{fig:fel-vs-r56-40nm}. The typical peak power reached at undulator exit is in the same range of a few tens of MW, and the optimum chicane strength still increases according to the transport line magnification.

\begin{figure}[h]
\includegraphics{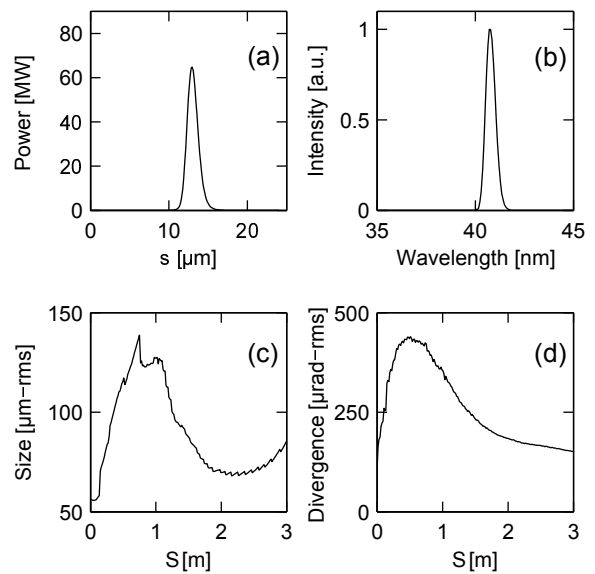}\\
\caption{\label{fig:fel-40nm} FEL (a) peak power longitudinal profile and (b) spectrum at the undulator exit. FEL (c) transverse size and (d) divergence along the undulator. FEL radiation at 40~nm in optimum case of Figure~\ref{fig:fel-vs-r56-40nm} ($r_{11}$=$r_{33}$=15 and $r_{56}$=0.25~mm). Beam transport simulation in the nonlinear case with BETA and sympletic 6D tracking without collective effects using the S2I-CM optics. FEL simulation with GENESIS.}
\end{figure}

The optimum performance is obtained with a magnification of 15 and a chicane strength of $r_{56}$=0.25~mm. In this case, as illustrated in Figure~\ref{fig:fel-40nm}, the peak power at the undulator exit is 70~MW for a pulse duration of 5~fs-fwhm (1.6~$\mu$m-fwhm), corresponding to a pulse energy of 0.35~$\mu$J.
The final FEL spectral width is 0.7~nm-fwhm, i.e. less than 2$\%$, and the FEL spot size is 80~$\mu$m-rms at the undulator exit for a divergence below 150~$\mu$rad-rms.

The required increase of energy, from 200 to 400~MeV, leaves the FEL sensitivity to the LPA parameters unchanged but relaxes all collective effects. The space charge is indeed directly related to the inverse of the energy, while the CSR is related only to the relative energy spread. 

The shortening to 40~nm of the radiation wavelength reduces of course the slippage length along the bunch, but the efficiency of the chromatic matching is maintained as shown in Figure~\ref{fig:fel-vs-r56-40nm}.

%%%%%%%%%%%%%%%%%%%%%%%%%%%%%%%%%%%%%%%%%%%%%%%%%%%%%%%%%%%%%%%%%%%%%%%%%%%%%%%%%%%%%%%%
\section{Conclusion}
%%%%%%%%%%%%%%%%%%%%%%%%%%%%%%%%%%%%%%%%%%%%%%%%%%%%%%%%%%%%%%%%%%%%%%%%%%%%%%%%%%%%%%%%

This paper presents a detailed study of an LPA based FEL at 200~nm. With typical sate-of-the-art LPA beam parameters, FEL amplification cannot be achieved unless a transfer line is used to manipulate this beam and mitigate its initial large energy spread and divergence. The appropriate line relies on three PMQs, four EMQs and one in-vacuum undulator to set a chromatic matching optics. 
To set the optics, one should begin by defining the desired magnification, which will fix the undulator taper, and then set the chicane strength which will remain the only variable parameter to optimize the FEL.
The robustness of the electron beam quality and consequent FEL performance, was analyzed first as a function of the LPA beam parameters and inherent jitters. Because of  chromatic effects, via emittance growth or orbits deviation, the sensitivity to the LPA beam charge and divergence was found to be extremely critical.
However the considered optics allows for an interesting robustness to the LPA pointing, which is a significant advantage considering the present stability of the LPAs. 
Sensitivity and robustness were then analyzed as a function of the magnetic element misalignments and/or defaults. The most critical point was found to be the alignment of the first set of PMQs, on which relies the initial refocussing of the LPA beam. But this sensitivity can be significantly mitigated applying a BPAC.
We also found that orbit correction enables to compensate for magnetic field errors even if this might be difficult to achieve in practice, together with the taper optimization, due to the LPA beam fluctuations.

Final simulations presented in the case of a 40~nm LPA based FEL allow interesting perpsectives of extended applications.

\begin{acknowledgments}
This work was partially supported by the European Research Council for the Advanced Grants COXINEL (340015). The authors would like to thank members of the Accelerator and Engineering Division and of the Experimental Division of SOLEIL.
The authors would also like to acknowledge V. Malka and his team from Laboratoire d'Optique Appliquée (Palaiseau) for fruitful discussions on the parameters set.
\end{acknowledgments}

% Create the reference section using BibTeX:
\bibliographystyle{unsrt}
\bibliography{Biblio-COX-transport}

\end{document}